\documentclass{osa-article}

\journal{osajournal}

\setcopyright{}

\usepackage{mathtools}
\usepackage[version=4]{mhchem}

\begin{document}

\title{Enabling high repetition rate nonlinear THz science with a kilowatt-class sub-100 fs laser source}


\author{Patrick L. Kramer,\authormark{1,*} Matthew Windeler,\authormark{1,2} Katalin Mecseki,\authormark{1} Elio G. Champenois,\authormark{1} Matthias C. Hoffmann,\authormark{1} and Franz Tavella\authormark{1}}
\address{\authormark{1}SLAC National Accelerator Laboratory, 2575 Sand Hill Road, Menlo Park, CA 94025,USA\\
\authormark{2}Queen's University, Kingston, Ontario, K7L 3N6, Canada\\}

\email{\authormark{*}pkramer@slac.stanford.edu} 



\begin{abstract}
Manipulating the atomic and electronic structure of matter with strong terahertz (THz) fields while probing the response with ultrafast pulses at x-ray free electron lasers (FELs) has offered unique insights into a multitude of physical phenomena in solid state and atomic physics. Recent upgrades of x-ray FEL facilities are pushing to much higher repetition rates, enabling unprecedented signal to noise for pump probe experiments. This requires the development of suitable THz pump sources that are able to deliver intense pulses at compatible repetition rates.  Here we present a high power laser-driven THz source based on optical rectification in \ce{LiNbO3} using tilted pulse front pumping. Our source is driven by a kilowatt-level Yb:YAG amplifier system operating at 100~kHz repetition rate and employing nonlinear spectral broadening and recompression to achieve sub-100~fs pulses at 1030~nm wavelength. We demonstrate a maximum of 144~mW average THz power (1.44~$\mu$J pulse energy), consisting of single-cycle pulses centered at 0.6~THz with a peak electric field strength exceeding 150~kV/cm. These high field pulses open up a range of possibilities for nonlinear time-resolved experiments with x-ray probing at unprecedented rates.
\end{abstract}


\section{Introduction}

Laser-based generation of high-field single- and multi-cycle terahertz (THz) pulses offers a powerful and widely accessible means of coherently driving atomic displacements or directly exciting specific low frequency modes to better understand collective mode (e.g., lattice) and electron dynamics in a wide variety of materials~\cite{Jepsen_2011, Hoffmann_2011, Hebling:08}. While accelerator-based sources can often provide higher pulse energies and a wider frequency tuning range, they are generally accompanied by high construction and operating costs, along with limited accessibility to researchers~\cite{green2016high}. Thus, THz generation with ultrafast laser pumping is ideal for many time-resolved experiments that require flexibility in delivering the THz pulses to the sample and synchronization with laser pulses at different frequencies (that are commonly generated from the same optical laser source) in pump-probe type experiments~\cite{Elsaesser_book_2019}. 

Terahertz pulses are widely used at x-ray free electron laser facilities for pump probe experiments~\cite{Kubacka2014,kozina2019,Gray2018} and in x-ray pulse characterization~\cite{Hoffmann_2018}. The future direction of large scale facilities is higher repetition rate (LCLS-II~\cite{abbamonte2015new}, FLASH-II~\cite{faatz2017flash}, European XFEL~\cite{cartlidge2016european}), in order to meet the needs of more complex but highly information-rich experiments, such as time-resolved resonant inelastic x-ray scattering (RIXS)~\cite{wernet2015orbital}, that will require many more detected photons during a measurement run than the lower repetition rate sources available so far have been able to provide. The higher repetition rate triggers challenges for the development of optical pump-probe lasers~\cite{mecseki2019high,pergament2014high}, and opportunities to develop high power sources in the entire spectral range from UV to THz, used for optical pump-probe experiments. 

This work explores the generation of THz pulses at 100~kHz repetition rate, gives direction for power scaling, and proposes routes to obtaining higher THz conversion efficiencies. We use the nonlinear compression technique to shorten pulses from a sub-picosecond Yb:YAG amplifier system to pump the THz source. Early work on nonlinear compression reported propagation in optical fiber for spectral broadening~\cite{fork1987compression}. Different methods have since emerged with a wide range of properties. Spectral broadening in a multi-pass cell combines the optimal beam quality of a hollow core gas-filled capillary output, with the simplicity of a single filament generated in gas~\cite{hanna2017nonlinear}. At the same time, this method clears the way for applications at higher energies and higher average powers. Recent advances in nonlinear compression with multi-pass cells show remarkable progress at tens of MHz repetition rate, using bulk elements in the focal plane~\cite{weitenberg2017multi,fritsch2018all}. Higher energy spectral broadening has been demonstrated in a Herriott cell arrangement~\cite{lavenu2018nonlinear,ueffing2018nonlinear}, with 18~mJ pulses reported by Kaumann {\em et al.} in Ref.~\cite{kaumanns2018multipass}. This work targets nonlinear compression factors (NCF) close to 10, with an average input power of 0.84~kW (100~kHz) and an input pulse duration of $\sim 700$~fs. The compressed output (70-100~fs) is used for THz generation experiments.

\section{Multi-pass cell setup and THz generation methodology}

\begin{figure}[tb]
\centering\includegraphics[width=9cm]{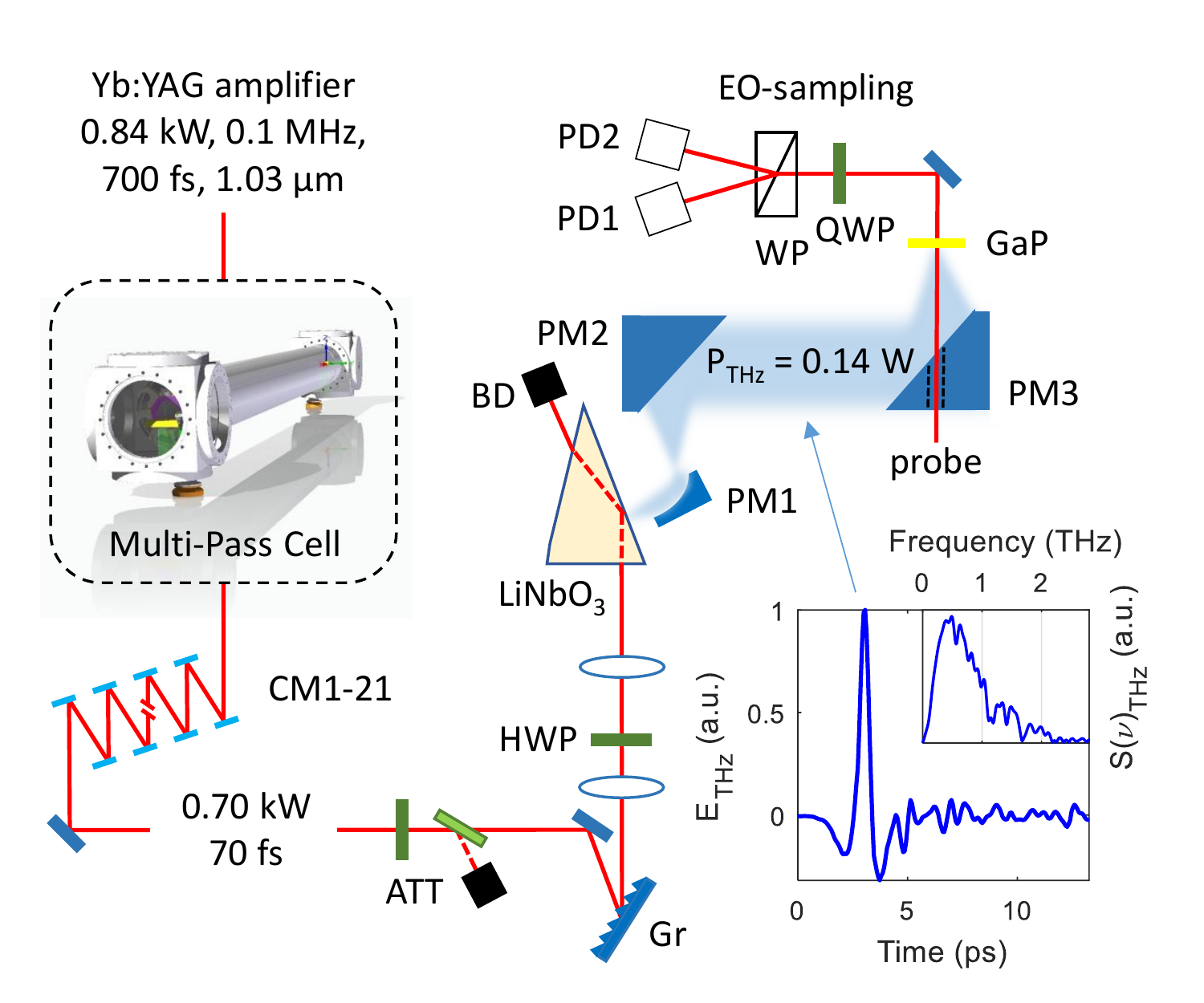}
\caption{Nonlinear compression and THz generation setup. The arrangement consists of a high power Yb:YAG amplifier system, a gas-filled multipass cell, a chirped mirror compressor, and an optical rectification setup with electro-optic sampling diagnostic. The inset shows a typical measurement of the THz electric field and its corresponding spectrum. Note: the highest Herriott cell output parameters are stated. Different settings have been used to optimize THz generation. CM - chirped mirrors, ATT - optical attenuator, Gr - reflection grating, HWP - half-wave plate, BD - beam dump, PM - parabolic mirror, QWP - quarter-wave plate, WP - Wollaston prism, PD - photodiode.}
\label{FIG1}
\end{figure}

A spectrally broadened and compressed Yb:YAG laser amplifier is used to drive a single-cycle THz pulse source based on optical rectification (OR) with tilted-pulse-front (TPF) pumping in lithium niobate (\ce{LiNbO3}). The experimental setup is illustrated in Fig.~\ref{FIG1}. 

\subsection{Yb:YAG amplifier system}

 The sub-picosecond 100~kHz Yb:YAG amplifier is based on Innoslab technology~\cite{russbueldt2010compact} (Amphos GmbH). The amplifier is primarily used as the pump for an optical parametric chirped-pulse amplifier; the parameters have been described previously~\cite{mecseki2019high,mecseki2019dual}. The output power was increased with improved thermal contacting of higher quality crystals in the final 2-pass amplifier~\cite{mecseki2019dual}. This reduces spatial filtering losses, required to improve output beam quality. 

\begin{figure}[tb]
\centering\includegraphics[width=10cm]{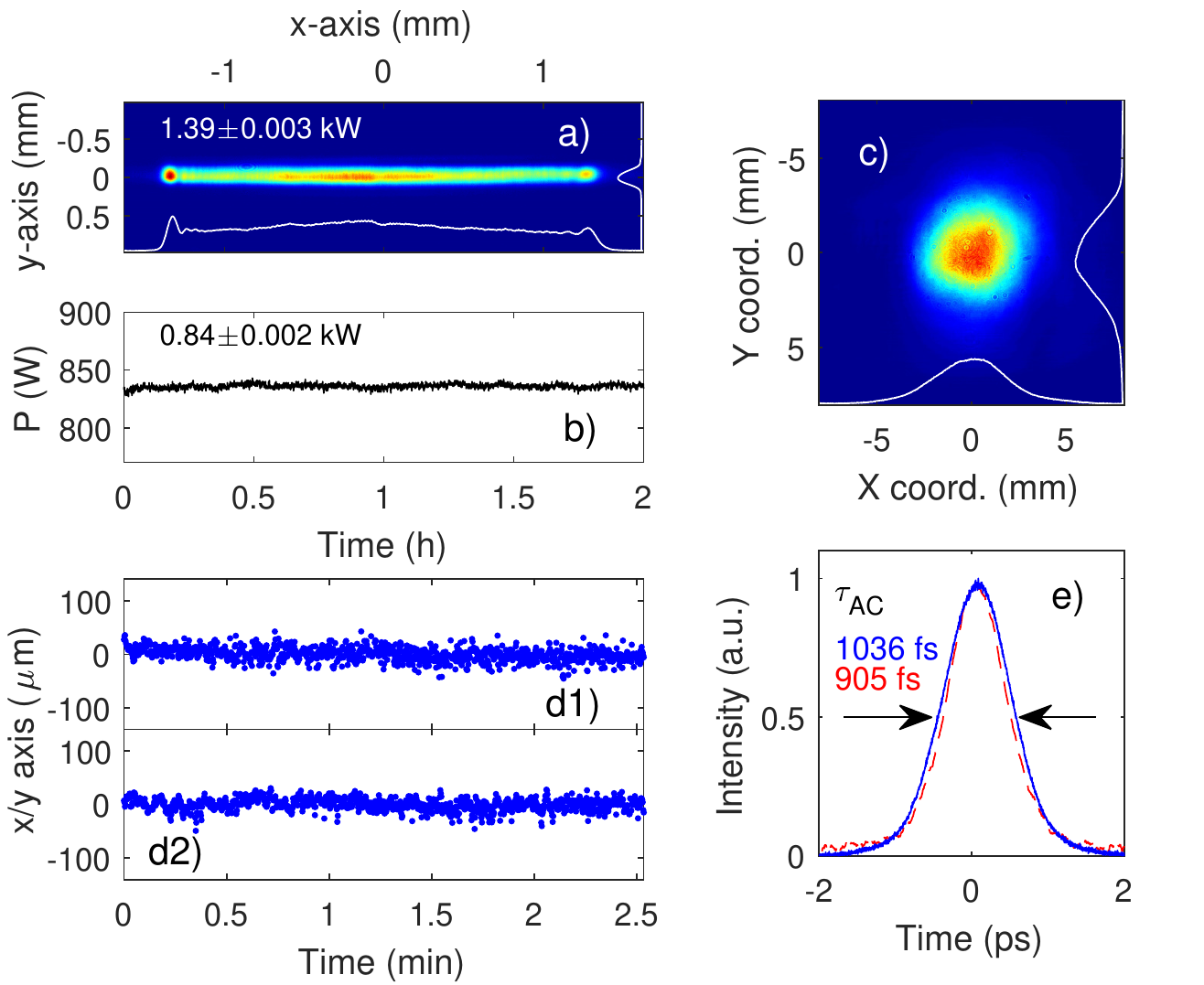}
\caption{Yb:YAG Innoslab amplifier system. (a) Imaged beam profile after 2-pass Yb:YAG kilowatt amplifier. (b) Power measurement after spatial filtering and compression. (c) Spatially filtered beam after equalization of beam aspect ratio and compression. (d) Beam pointing stability measurement. (d1) $x$-axis, (d2) $y$-axis. (e) Auto-correlation measurement of the Yb:YAG pump beam with evacuated MPC (blue) and with 1.1~bar Argon at low power (dashed red, 7~W). The deconvoluted pulse durations are 732.6~fs (blue) and 639.9~fs (dashed red), assuming a Gaussian temporal distribution.}
\label{FIG5}
\end{figure}

Figure~\ref{FIG5}(a) shows an image of the 1.39~kW amplified beam, recorded after the final 2-pass Innoslab amplifier. Two cylindrical telescopes are used to re-shape the beam to equal aspect ratio. The beam is Gaussian in the $y$- axis, but shows multi-mode structure in the $x$-axis, which is attributed to thermally-induced phase differences in the amplifier crystal. These higher frequency components are spatially filtered before compression. The compressed output power is 836.0$\pm$2.0~W, as shown in Fig.~\ref{FIG5}(b). A measurement of the spatial beam profile after compression is shown in Fig.~\ref{FIG5}(c). The beam size is 7.7~mm in diameter at $1/e^{2}$. The beam is demagnified with a 1.5:1 telescope before entering the multi-pass cell. 

One operational challenge is an inherent thermal beam drift on a timescale of tens to hundreds of milliseconds. A beam stabilization system (MRC Systems) was implemented in the compressor setup to stabilize beam pointing. The system consists of two actuators and two 4-quadrant diodes. Both actuators are placed before the spatial filter and compressor, along with one of the 4-quadrant detectors. The second detector is placed close to the application, in this case the multi-pass cell. The corrected beam pointing trajectories are shown in Fig.~\ref{FIG5}(d1) and (d2). The $x$- and $y$-axis deviations are 12.9~$\mu$m and 11.4~$\mu$m, root mean square (rms), respectively. A second beam stabilization system is used to stabilize the beam path through the multi-pass cell and to the THz generation setup. The first detector is placed at the multi-pass cell output. The second 4-quadrant detector is placed at a virtual plane in front of the THz generation setup. The use of beam stabilization systems enables the highly reproducible day-to-day performance of the spectral broadening setup. The corrected beam pointing drifts at the THz setup ($x$ and $y$) are 13.2~$\mu$m and 23.6~$\mu$m, rms. 

The pulse duration measurement in Fig.~\ref{FIG5}(e) shows an auto-correlation measurement taken at low power with evacuated multi-pass cell at its output (blue line, $\tau_{p}$ = 732.6~fs assuming a Gaussian temporal distribution), and with 1.1~ bar Argon in the MPC, after the chirped compressor (red line, $\tau_{p}$ = 639.9~fs). The pulse duration at lower Argon gas pressures, below 700~mbar, is 680~fs (not shown).

\subsection{MPC design}

The spectral broadening is performed in a Herriott cell that can be configured for a maximum of 14 focalizations. Single mirrors are placed in a circular arrangement to mimic the beam path of a traditional Herriott cell. The near-concentric cavity mirror arrangements are placed inside two 6-way cubes set 2~m apart (see rendering in Fig.~\ref{FIG1}). The input and output windows are AR-coated 3mm thick UV-fused silica flats. The initial focusing mirror and the last re-collimating mirror have a radius of curvature (RoC) of 2~m. All other remaining cell mirrors have an RoC of 1~m. The cell is operated with a fixed chirped mirror compressor with total dispersion of -21500 fs$^{2}$. Cell parameters can be varied mainly with the pressure of the broadening medium (Argon) and the input peak power. The Herriott cell output power can be varied with an optical attenuator. THz generation has been operated with up to 384~W of average power, where thermally-induced absorption in the \ce{LiNbO3} medium severely limits further conversion. 

Insight into the operation of the MPC over a wide range of parameters was provided by numerical calculations. The spectral broadening process was modeled using a 2D numerical split-step Fourier simulation to solve the nonlinear Schr\"odinger equation \cite{Boyd_3ed}. Further detail on the 2D simulations and the extension to a full 3D model will be provided in a forthcoming publication.

\subsection{THz generation and characterization}

Following the variable attenuator, the spectrally broadened and compressed pulse emerges horizontally polarized and is directed by a beamsplitter into two paths: a pump arm (99\% power) for generation of intense terahertz pulses and a probe arm (1\% power) for electro-optic sampling (EOS) of the resulting THz waveforms. The THz generation pump path is illustrated in Fig.~\ref{FIG1}; it follows the highly successful scheme for TPF pumping, described in detail previously~\cite{hebling2002, Hebling:08, Hoffmann_2011}, to achieve noncollinear velocity matching between the femtosecond pump pulse and the terahertz wave generated through OR in \ce{LiNbO3}. 

Both the pump and THz waves are polarized along the \ce{LiNbO3} $z$-axis (i.e., both \emph{e}-polarized) to use the largest nonlinear coefficient $d_{33}$ for OR. The phase matching condition is satisfied for THz wave propagation at an angle $\gamma$ relative to the pump propagation direction, with $\gamma$ determined by the noncollinear velocity matching equation: $v_g(\omega_0) \cos \gamma = v_p(\Omega)$. Here $v_g(\omega_0)$ is the group velocity of the pump pulse centered at frequency $\omega_0$ and $v_p(\Omega)$ is the phase velocity of the terahertz pulse centered at $\Omega$~\cite{hebling2002}. For pump pulses centered at 1.03~$\mu$m and terahertz wave frequencies around 1~THz, \ce{LiNbO3} has a pump group index of $n_g = 2.22$ and a THz refractive index of $n_\text{THz} = 4.96$. The velocity matching condition can be equivalently written as $n_\text{THz} \cos \gamma = n_g$. For these parameters we obtain a value of $\gamma = 63.41^\circ$ for the necessary pulse front tilt angle inside the crystal. 

Experimentally, we introduce pulse front tilt to the pump laser source by diffraction from a grating structure. A 1600 lines/mm dielectric reflective diffraction grating at the $+1$ order~\cite{gratinghandbook} and a -1.89:1 telescope constructed with a pair of cylindrical lenses having focal lengths $+151$~mm and $+80$~mm are used to achieve the correct tilt angle. Between the first and second lens, a zero-order half wave plate rotates the pump polarization from horizontal (for maximum diffraction efficiency from the grating) to vertical. 

THz generation was evaluated using two different magnesium oxide (\ce{MgO}) doped stoichiometric lithium niobate (\ce{LiNbO3}) prisms. The first was supplied by Oxide Corporation, with angles $\alpha, \beta = 62 ^ \circ$ (isosceles prism design), a front face width of 10 mm, a height of 12.5 mm, and which has 1.3 mol \% \ce{MgO} content. The second was supplied by the Wigner Research Center for Physics, with angles $\alpha = 63 ^\circ$ and $\beta = 44 ^ \circ$ (Hebling prism design), a front face width of 10 mm, and a height of 25 mm. This crystal has lower \ce{MgO} doping, at 0.6 mol \%. Almost identical output was obtained with both (see results in Section 3.2), suggesting that ultimately the conversion is limited by other factors and both geometries and doping levels are effective. The crystals were water cooled via aluminum blocks placed in thermal contact on the top and bottom surfaces. A recirculating chiller set to $15^\circ$C provided cooling water through both blocks as a single loop. A thermocouple monitored the temperature of the upper block, providing a proxy for the increasing crystal temperature while ramping up pump power. The THz field is emitted from the \ce{LiNbO3} prism at the angle $\gamma$ from the pump input and is coupled out at approximately normal incidence in both prism geometries. 

Care needed to be taken to account for and dispose of all residual, high power, pump light exiting the \ce{LiNbO3} crystal. The majority of the unconverted pump light is coupled out of the polished third or ``back'' surface of the prism and is directed into a water-cooled beam dump. Back-reflected beams from the pump input, ``front,'' face were blocked with matte black coated aluminum foil to protect upstream optics and their mounts. Finally, with both the Hebling- and isosceles-design \ce{LiNbO3} prisms, some amount of residual pump energy exits from the THz out-coupling face after generally multiple reflections inside the crystal. The beam propagates nearly parallel to the emitted THz radiation and can be collected by the parabolic mirror imaging setup if it is not blocked. The isosceles-cut prism, while having a better collimated out-coupled pump from the back surface, was found to direct considerably more pump light into the THz path than the Hebling-cut prism did. The Hebling design is thus preferred in high average power setups due to the difficulty in separating THz and pump frequencies without additional loss.

THz radiation was separated from the residual pump with a dual-plastic filter, consisting of a 3.175~mm thick polytetrafluoroethylene (PTFE) sheet, followed by a black colored (highly absorptive for near IR) approximately 2~mm thick polypropylene (PP) sheet. The residual pump beam would quickly heat and begin to melt the PP filter on direct impact, but the PTFE sheet served to scatter and diffuse the beam, allowing the PP filter to absorb the pump radiation over a wider surface area. Ensuring the THz beam was free of pump radiation was particularly important during measurements of the THz average power scaling. A 2~mm thick, 50.8~mm diameter high-resistivity silicon wafer with a broadband high-reflective coating centered around 1030~nm was additionally investigated as a THz low pass filter. We found significant thermally-induced absorption at high pump powers which made it unsuitable for high power operation. 

Three metallic off-axis parabolic mirrors (PM) following the \ce{LiNbO3} crystal serve to expand, collimate, and then re-focus the THz beam at the the final (sample) position. The first is a $117^\circ$ angle PM with $f = 25.4$~mm and 25.4~mm diameter, placed approximately one focal length away from the output face of the prism for an intermediate THz focus. The remaining two have focal lengths of 177.8~mm and 76.2~mm, respectively, with 76.2~mm diameters and $90^\circ$ angles. The intermediate THz focus is imaged into the sample position with demagnification factor of approximately 2.3. The plastic low-pass filters were placed just prior to the final focusing PM. 

The THz power was measured at the focus using a calibrated thermopile sensor (model 3A-P-THz, Ophir). Spatial mode measurements of the focused THz spot were made with a pyroelectric camera (Pyrocam IIIHR, Ophir). The probe pulse was variably delayed and attenuated, set to vertical polarization, and focused through a hole in the third PM to overlap the focused THz beam at the sample position. EOS measurements were made in 110-oriented gallium phosphide (\ce{GaP}) crystals with thickness of 100~$\mu$m or 500~$\mu$m. The probe then passes through a quarter wave plate and Wollaston prism and intensities $I_1$ and $I_2$ are detected on a balanced photodiode (PD) pair (DET100A, Thorlabs). The difference signal, $I_1 - I_2$, is recorded with a lock-in amplifier (SRS 830, Stanford Research Systems) for the time-dependent EOS traces, and the normalized modulation $(I_1 - I_2)/(I_1 + I_2)$ was determined at probe delays for the THz field peak by averaging several PD signals with an oscilloscope (Tektronix TDS 3054C) for calculation of absolute electric field values~\cite{Planken:01}. 

\section{Experimental results and discussion}

\subsection{Spectral broadening of the Yb:YAG laser system}

The multi-pass cell (MPC) is operated within a pressure range of 0.5-2~bar Argon and up to 10~GW peak power. Previously published literature on MPC spectral broadening use a high number of passes with a low single-pass phase shift $\varphi < 1$~rad, with few exceptions, such as in Ref.~\cite{kaumanns2018multipass}, where a moderate single pass nonlinear phase shift is used ($\varphi \sim 1.2$~rad). We explore the possibility to use a reduced pass number with a higher nonlinear phase per pass. A compact setup with low number of passes improves stability, is safer to operate at kW average powers, and reduces setup complexity. 

\begin{figure}[tb]
\centering\includegraphics[width=11.5cm]{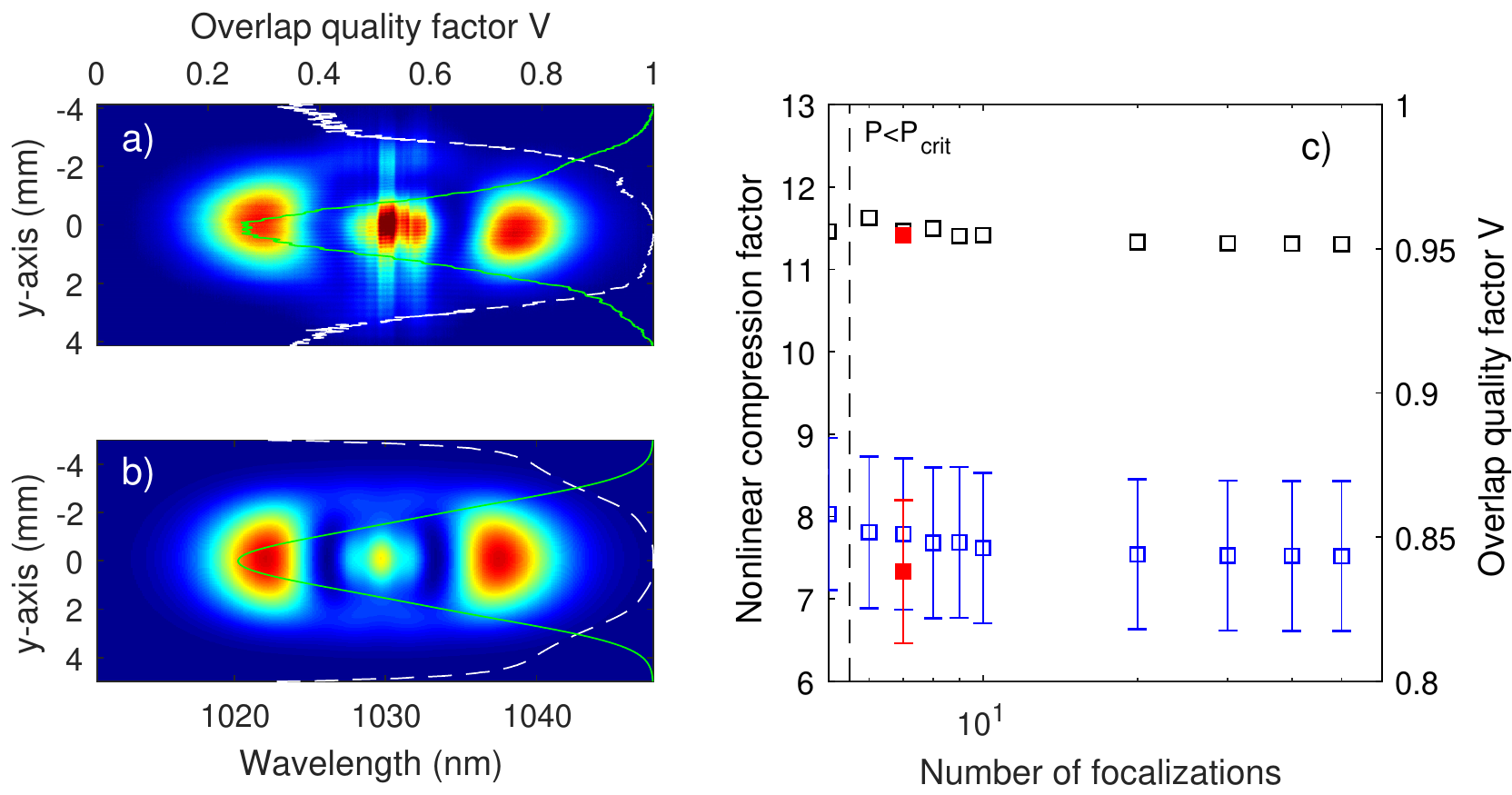}
\caption{Comparison of experimental and simulated spatio-spectral distributions of the MPC output. (a) Experimental data and (b) simulation with overlap quality factor $V$ (dotted white) and the integrated beam profile along the spectral axis (green). Note: the experimental data at 1.03~$\mu$m is saturated for better comparison to simulation. (c) Simulation of $V$ (black squares) and nonlinear compression factor (NCF, blue squares) versus number of focalizations (log-horizontal scale), keeping the NCF roughly constant. The critical power threshold is marked as a dotted black line. Red data point are values for $V$ and NCF for the experimental data in (a).}
\label{FIG14}
\end{figure}

\begin{figure}[tb]
\centering\includegraphics[width=13cm]{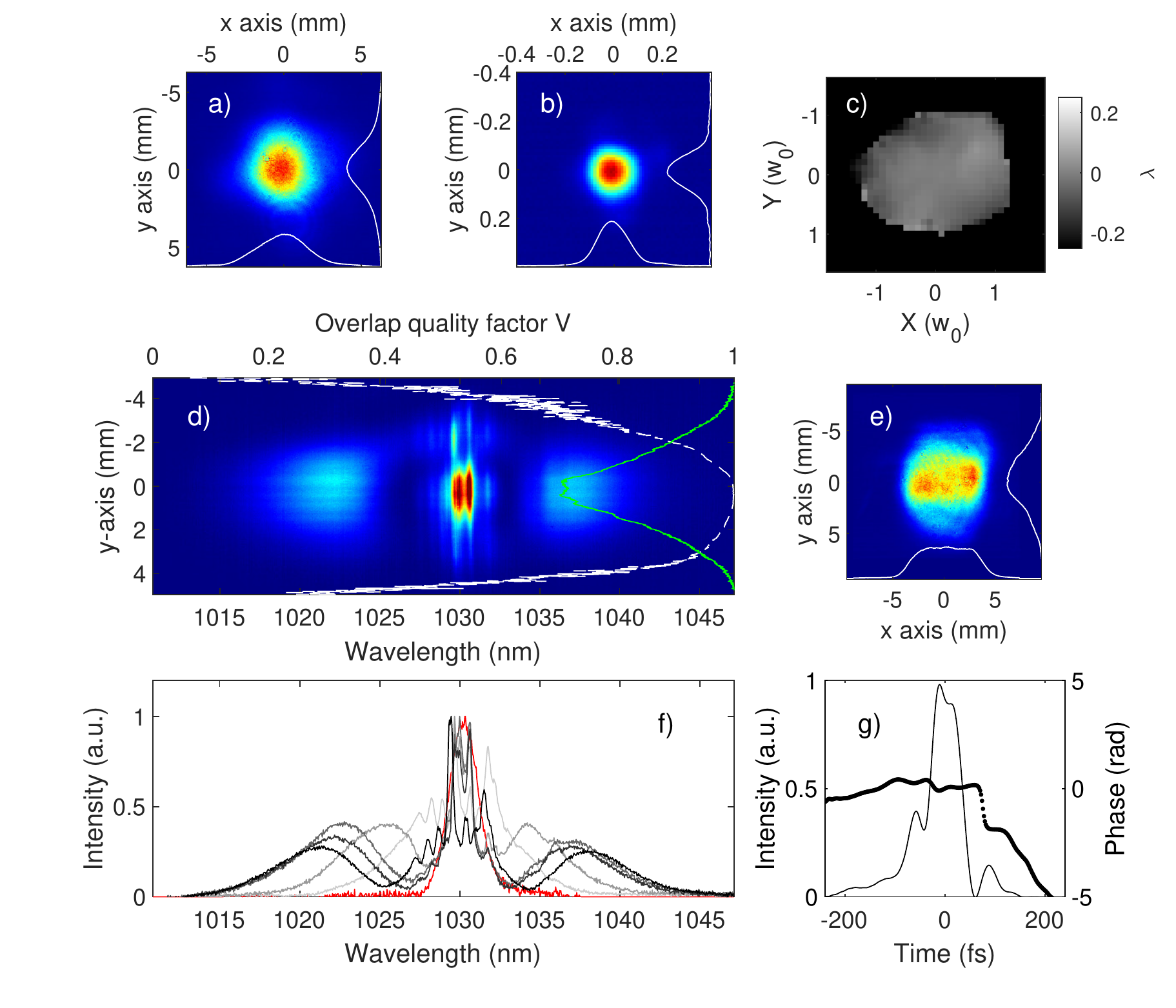}
\caption{Characteristics of MPC output at the highest average input power: (a) Spatial beam profile and (b) focus at the MPC output. (c) Wavefront measurement of the compressed output (note: axes normalized to beam waist $w_0$). (d) Spatio-spectral distribution of the output beam. Overlap quality factor (white), integrated intensity distribution (green). (e) Beam profile after compression with 0.74~kW power on chirped mirrors. 
(f) Ar gas cell pressure scan of output spectrum. (g) Measured intensity and phase distribution of the compressed pulses (FROG, black lines).}
\label{FIG6}
\end{figure}

The MPC spectral broadening effect is modelled in a 2D split-step Fourier simulation with rotational symmetry. The overlap quality factor, often used as a figure of merit for spectral beam uniformity, is defined as
\[
V = \frac{\left[ \int A(\lambda)A_{0}(\lambda)\,\mathrm{d}\lambda \right]^2}{\int A(\lambda)^2\,\mathrm{d}\lambda \times \int A_{0}(\lambda)^2 \, \mathrm{d}\lambda}.
\]
$A(\lambda)$ is the spectral amplitude and $A_{0}(\lambda)$ is a reference amplitude at the geometric beam center. The overlap quality factors of the system described in this paper and in previously published work (e.g. Ref.~\cite{kaumanns2018multipass,lavenu2018nonlinear}) are similar, with average values of $V > 0.95$ within the waist diameter of the beam. We use an additional metric, the normalized rms deviation (NRMSD) of the Fourier-limited pulse duration, defined as $\mathrm{NMRSD} = \mathrm{RMSD} / \langle\tau_{4\sigma}\rangle$, to capture the spectral quality across the beam. The average change of pulse duration across the beam is a figure of merit for temporal uniformity. Figure~\ref{FIG14}(a) and (b) show a comparison of measured (Acton 4-f spectrometer) and simulated spatio-spectral distribution of the MPC output for the highest input pulse energy and the highest nonlinear compression factor, $\mathrm{NCF} = \tau_\text{NC}/\tau_{0}$. Fig.~\ref{FIG14}(c) shows $V$, the NCF, and the NRMSD of $\tau_\text{NC}$ (error bars of the NCF) with increasing number of MPC passes (focalizations) in the simulation. The input power is 0.84~kW and the beam waist diameter is 5~mm (at $1/e^{2}$). The cavity mirrors have an RoC of 1~m and the cavity length is slightly detuned from its confocal arrangement. The input pulse duration of the Yb:YAG amplifier is 675~fs. With increasing number of focalizations, neither $V$ nor $\tau_{4\sigma}$ substantially improve. This lack of change is due to rather steep focalization in this compact arrangements (2~m cavity), at a relatively high peak power of approximately 10~GW. The MPC is therefore setup in a 7-fold focal arrangement, slightly below the critical power, and results are shown Fig~\ref{FIG6}. The average values for the overlap quality factor $V$ are 0.945 and 0.956 for experiment in Fig.~\ref{FIG14}(a) and simulation in Fig.~\ref{FIG14}(b), respectively. These values are presented in Fig.~\ref{FIG14}(c).  

The focal waist and beam size on the focusing mirrors are set close to the cavity eigenmode size (Fig.~\ref{FIG6}(a) and (b)) for near-concentric resonator condition. The waist diameter of the resonator has a solutions for the fundamental $\mathrm{TEM}_{00}$ wave, given in Ref.~\cite{siegman1986lasers}. The expected $1/e^{2}$ beam diameter on the mirror surface is 4.89~mm. A measurement is shown in Fig.~\ref{FIG6}(a) with 4.97~mm and 4.81~mm in the $x$- and $y$-axes, respectively. The waist diameter of the resonator has a solution of 268~$\mu$m for a deviation $\Delta L \approx 6$~mm. A focal measurement is shown in Fig.~\ref{FIG6}(b), with waist diameters 272~$\mu$m and 279~$\mu$m in $x$ and $y$. The optical wavefront variation across the beam profile is $0.15 \lambda$ peak to valley (Fig.~\ref{FIG6}(c)). A typical spatio-spectral measurement of the output beam at 1~bar Ar is shown in Fig.~\ref{FIG6}(d), along with the overlap quality factor $V$ (dotted white line) and the integrated intensity along the spectral dimension (green). The compressed output of the Yb:YAG amplifier system is 836.0$\pm$1.9~W. The highest output from the Herriott cell is 736.5$\pm$5~W and is compressed to $\sim 0.7$~kW. The Herriott cell transmission is 88\% and the compressor transmission is 94\%. Additional transport and optical attenuator losses are 3\%. The highest power used in the THz setup is $384.2 \pm 1.7$~W, with a clear conversion efficiency roll-over observed at approximately 300~W. The spectrum of the broadened output pulses are shown in Fig.~\ref{FIG6}(f) as a function of Ar gas pressure. The nonlinear compression factor is approximately 10. A typical pulse duration measurement with Frequency Resolved Optical Gating (FROG) is shown in Fig.~\ref{FIG6}(g) (Swamp Optics, Grenouille. Pulse intensity: black; temporal phase: dotted black). The pulse duration is 70~fs at FWHM.

No thermal effects have been observed during operation of the multi-pass cell. The optics used in the multi-pass cell are multilayer spherical mirrors (HR 980-1080~nm) and AR coated windows. The chirped mirror stacks however absorb a quantity of power, and heating of these mirrors was observed. The peak temperature at the center of the beam scales approximately with 10.9~mK/W (e.g. $27.5^{\circ}$C at 440~W). At average powers exceeding 500~W, cumulative thermal lensing effects from the large number of chirped mirrors reshape the beam profile,  from a Gaussian to a flat top, during propagation to the THz setup (Fig.~\ref{FIG6}e). We have therefore limited the power for long term operation through the compressor to avoid damage to the chirped mirrors. While performing THz experiments we have used average power just below 400~W, with a cell pressure of $P_\text{Ar}$ of 1.6~bar to attain a comparable NCF as at full power, and a comparable pulse duration below 100~fs. The average power of approximately 400~W is sufficiently high to reach past thermal limitations in the THz setup. The MPC output characteristics optimized for THz generation are shown in Appendix A, Fig.~\ref{FIG7:params}. The thermal lensing from the chirped mirrors is essentially eliminated (see beam profile in Fig.\ref{FIG7:params}(a). A next generation setup will include chirped mirrors with larger aperture and lower GDD/bounce to avoid excessive power absorption.

\subsection{THz generation}

\begin{figure}[tb]
    \centering
    \includegraphics[width=12.5cm]{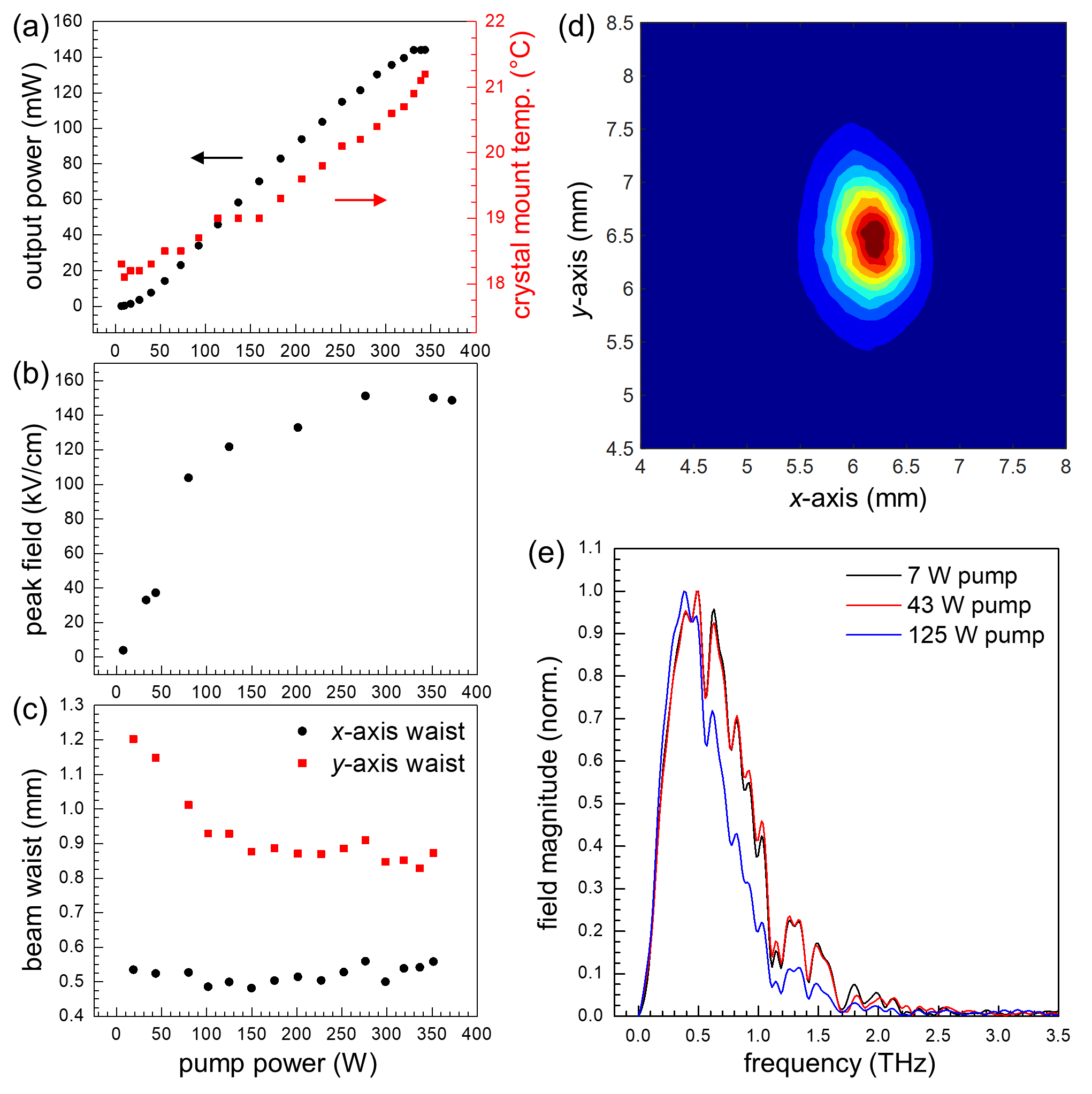}
    \caption{Characterization of high average power THz output. Parts (a) through (c) show variation of THz properties with incident pump power. (a) Steady-state output power (black, left axis) and crystal assembly temperature measured on top of cooling block (red, right axis). (b) Peak electric field determined from EOS in \ce{GaP}. (c) $1/e^2$ Gaussian beam waists at the THz focus. Part (d) displays a representative THz spatial mode at the focus, recorded at 352~W pump power. In (c) and (d), the $x$-axis corresponds to the vertical direction, while $y$ is horizontal, with respect to the \ce{LiNbO3} crystal. (e) THz electric field magnitude spectra from EOS for low to moderate pump powers (measured in a 500~$\mu$m thick \ce{GaP} crystal), showing the spectral narrowing due to increased THz absorption at high incident power.}
    \label{FigTHz1}
\end{figure}

Using the experimental arrangement illustrated in Fig.~\ref{FIG1} and described in Section 2.2, THz radiation was generated and the output characterized as a function of input pump power. Fig.~\ref{FigTHz1}(a) displays the average THz power in black (left axis), determined at the final focus position for input pump power between 7~W and about 375~W, at which point heating of the \ce{LiNbO3} crystal dominates and no further increases in THz output are observed. The temperature of the water-cooled \ce{LiNbO3} mount (a simple proxy for the crystal temperature) is also shown in Fig.~\ref{FigTHz1}(a) in red (right axis), which clearly tracks the increasing pump power. Average THz power measurements were recorded once the temperature reading reached steady-state. The crystal heating with each power increase resulted in an average of 4\% decrease in steady-state THz power compared to the immediate reading before thermalization.

The maximum THz average power observed was 144~mW (1.44~$\mu$J per pulse) at a pump power of 344~W, which represents, to the best of our knowledge, the highest average power single-cycle THz pulsed source reported to date. At this record power, the energy conversion efficiency is about $4.2 \times 10^{-4}$, corresponding to a photon efficiency of approximately 19\%. These values are comparable to previous reports using similar pump pulse energies~\cite{Hoffmann_2011}. 

The residual pump light coupled out of the \ce{LiNbO3} prism exhibited a clear red-shifted shoulder when the conditions were optimized for efficient THz generation (Appendix B, Fig.~\ref{FigTHzS2}), showing that the high photon conversion is the result of several cascaded steps. Not all of the generated THz light is coupled out of the crystal, however, as quantitative analysis of the  pump red-shift~\cite{Yeh_2007} computes a photon conversion efficiency of 246\% inside the crystal, with an external conversion efficiency of 137\% considering the Fresnel reflection loss. Unsurprisingly, most of the generated THz photons are lost to absorption throughout the \ce{LiNbO3} crystal interior (Section 3.3) or not collected by the exterior imaging optics. 

Considering both the cylindrical pump imaging optics and anamorphic magnification due to the angle of diffraction from the grating in estimating the illuminated spot dimensions at the crystal, the pump fluence was calculated as $11.4 ~ \mathrm{mJ/cm^2}$ for this maximum THz output. While there is precedent for improved conversion in \ce{LiNbO3} using even greater pump fluence values with a smaller pump spot, the thermal gradients induced by the high average intensity (see below) limit us to the present maximum fluence.  

The electric field temporal profiles and resulting spectra of the THz pulses were measured with EOS, using 110-oriented \ce{GaP} crystals of either 100~$\mu$m or 500~$\mu$m thickness, selected to keep the electro-optic modulation below 30\% to avoid saturation~\cite{Planken:01} without requiring attenuation of the THz field. A typical time trace of the electric field generated at about 35 W input pump power and the resulting magnitude spectrum  are shown  in Fig.~\ref{FIG1}. The shape is a clear single cycle pulse, characteristic of optical rectification in \ce{LiNbO3}. All of the THz generation experiments were conducted in ambient air; atmospheric water absorption is responsible for the sharp absorption lines in the spectrum and the corresponding oscillations following the main pulse in the time domain data. 

The peak electric field at the THz focus is displayed as a function of pump power in Fig.~\ref{FigTHz1}(b). The field values plateau at lower power compared to the THz pulse energy, reaching a maximum of 151~kV/cm peak field at an input power of 276~W. Electric field values at the focus depend not only on the THz pulse energy, but also the spectral content and beam profile at the \ce{LiNbO3} crystal, which is expanded then focused into the detection crystal. 

The dependence of $x$-axis and $y$-axis THz beam waist values ($w$, the $1/e^2$ intensity half width) on the incident pump power is displayed in Fig.~\ref{FigTHz1}(c). A typical image of the focused THz spot with 352~W pump power appears in Fig.~\ref{FigTHz1}(d). The $y$-axis waist (corresponding to the horizontal direction in the crystal) shows a strong trend of decreasing with increasing pump power, while the $x$-axis waist (the vertical direction in the crystal) does not change appreciably. Above 80~W pump power, the spot size no longer varies significantly, with average values $w_y = 0.89 \pm 0.05$~mm and $w_x = 0.52 \pm 0.03$~mm. The horizontal beam waist ($y$-axis) has been observed to vary much more significantly than the vertical ($x$-axis) in previous studies, where the narrowing with more efficient THz generation was attributed to nonlinear propagation effects in the \ce{LiNbO3} crystal~\cite{Lombosi_2015}. It could be expected that thermal lensing may also contribute, as the high average pump power leads to a notably inhomogeneous temperature distribution in the crystal (Section 3.3), but the relative invariance of the vertical beam waist suggests this is at most a minor contributor. 

\begin{figure}[tb]
    \centering
    \includegraphics[width=12.5cm]{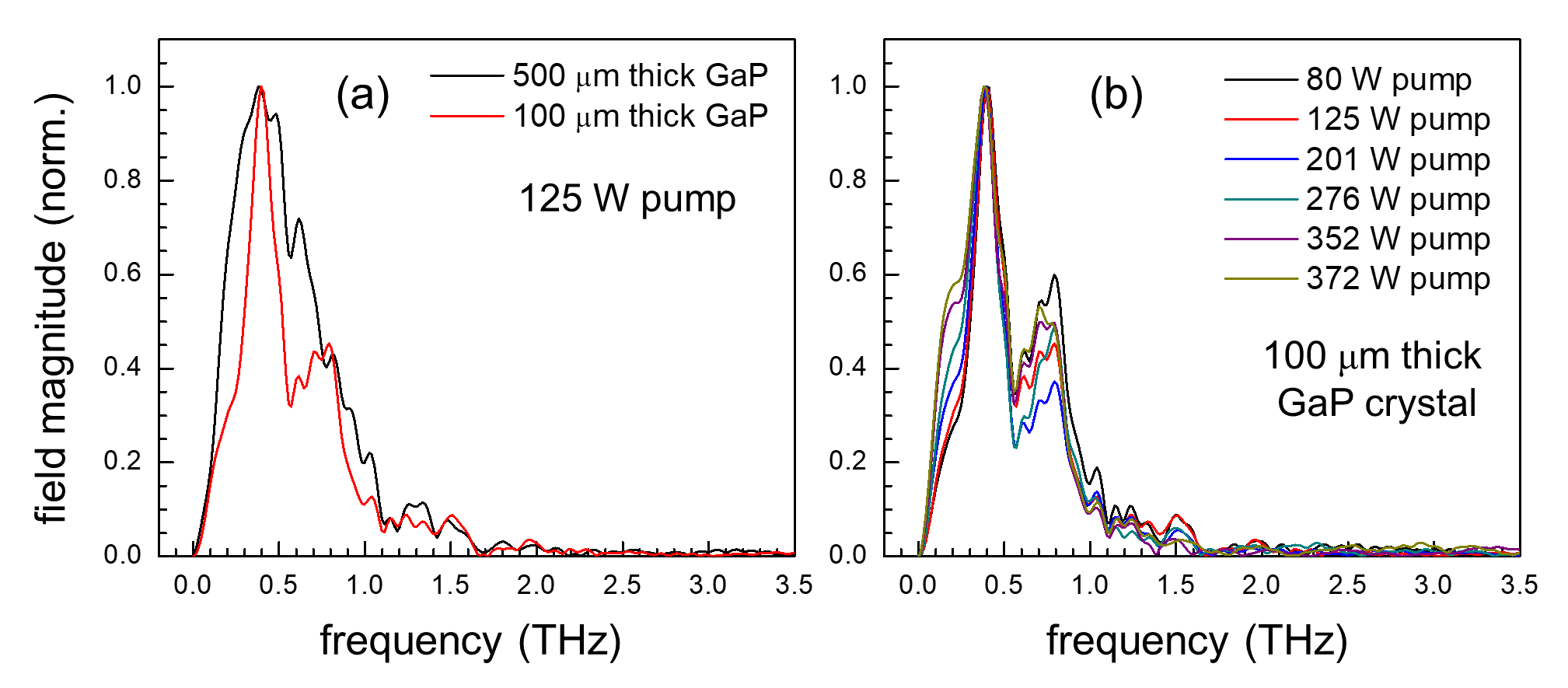}
    \caption{Electric field magnitude spectra from EOS measurements of the generated THz waveforms. (a) Comparison of spectra determined in 500~$\mu$m and 100~$\mu$m thick \ce{GaP} crystals at constant 125~W pump power. (b) Dependence of THz spectrum on pump power, with 100~$\mu$m thick \ce{GaP} crystal used throughout to avoid saturating the electro-optic response at the high THz fields generated above 125~W.}
    \label{FigTHzS3}
\end{figure}

Electric field magnitude spectra from EOS (normalized) are given for input pump power up to 125~W in Fig.~\ref{FigTHz1}(e). As the input pump power increases above 43~W, we see a significant drop in field amplitude at higher THz frequencies. This is due to the increased THz absorption, in particular for higher frequencies, as the \ce{LiNbO3} crystal increases in temperature~\cite{hebling2004,Huang:13}. Magnitude spectra for higher pump power values were recorded in a thinner (100~$\mu$m) \ce{GaP} crystal and are displayed in Fig.~\ref{FigTHzS3}. In Fig.~\ref{FigTHzS3} (a), the the 100~$\mu$m and 500~$\mu$m thick EOS detection crystals are compared for the same pump power. In the thinner crystal, a reflection of the THz pulse from the crystal surfaces appears as a delayed copy in the time window, which then interferes destructively in the frequency domain with the major fraction of the THz pulse whose detection is desired. The reflected THz pulse appears much later in time for the 500~$\mu$m \ce{GaP} crystal, outside the time window we scanned within. This interference is responsible for the dips in magnitude that appear in the trace with a 100~$\mu$m thick crystal as compared to 500~$\mu$m. 

In Fig.~\ref{FigTHzS3} (b), THz electric field magnitude spectra are displayed for input pump powers between 80~W and 372~W. The most significant trend is a gradual increase in relative magnitude for frequencies below 0.3~THz with increasing pump power. The change in higher-frequency components (above 0.5~THz) is much less for pump power levels above 125~W than the differences observed at lower power levels in Fig.~\ref{FigTHz1}(e). 

The relative loss of high frequency THz components at higher pump powers is most likely responsible for the saturation in peak field values shown in Fig.~\ref{FigTHz1}(b). While heating of the nonlinear crystal affects the THz output spectrum, conversion efficiency, and attainable peak electric field, it is important to note that these changes are entirely reversible upon allowing the \ce{LiNbO3} crystal to cool, with no discernible damage or alterations observed even after extended periods of THz generation at high pump intensity.

The spectrally broadened Yb:YAG kW-scale laser system was found to be an excellent pump source for THz generation by optical rectification in \ce{LiNbO3}, with highly repeatable day-to-day operation. If the crystal heating can be minimized while keeping the medium well below ambient temperature~\cite{hebling2004,Huang:13}, it is clear that the THz output could be scaled significantly higher due to both the better efficiency of OR and ability to use the full available pump pulse energy. 

\subsection{Thermal response of THz crystal}

\begin{figure}[tb]
\centering\includegraphics[width=7.5cm]{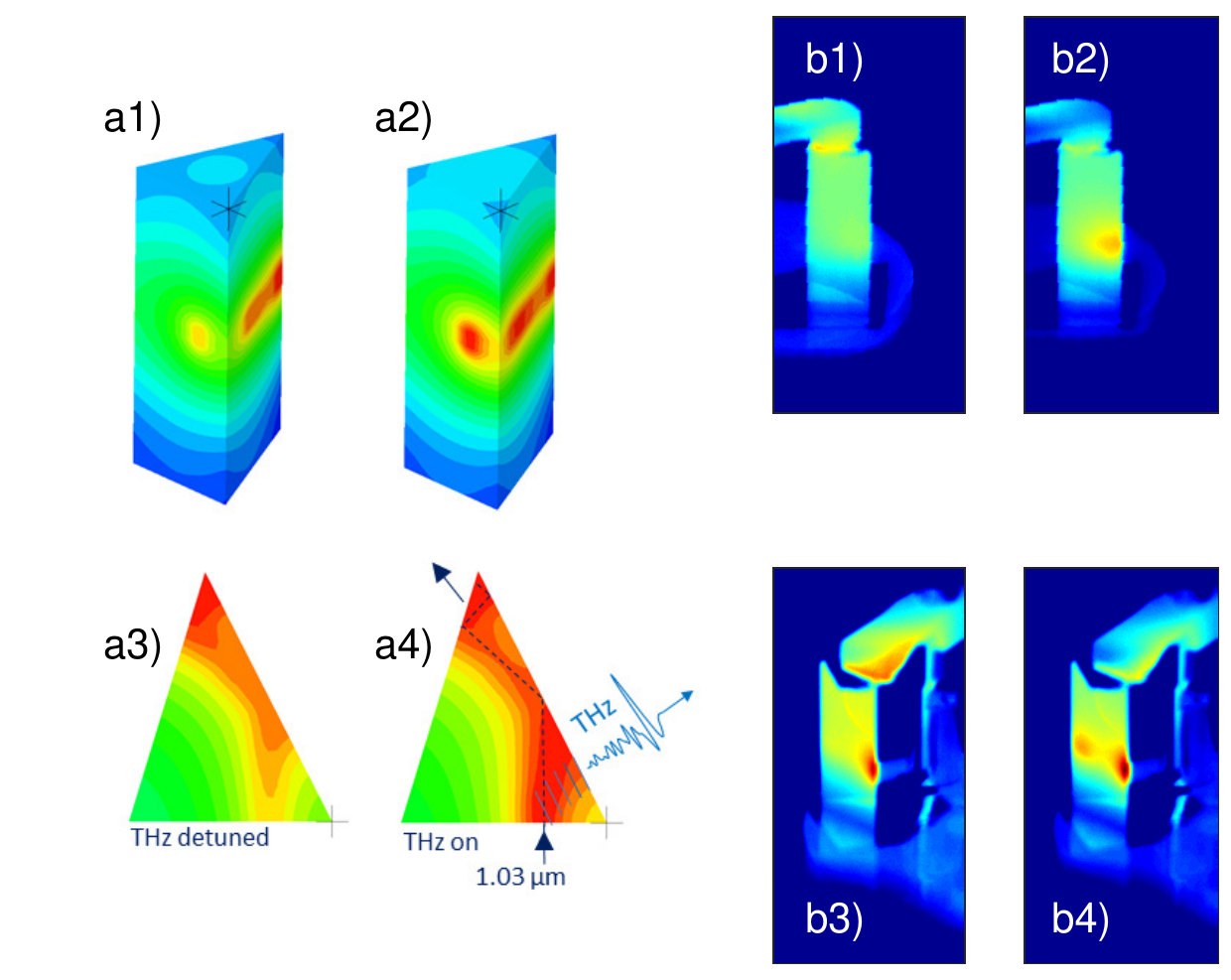}
\caption{Thermal images and simulation of \ce{LiNbO3} crystal temperature distribution. (a1), (a2) Isometric view and (a3), (a4) cross-section view of a Hebling-cut prism. The simulation shows pump absorption only (left, a1 and a3) and pump and THz absorption (right, a2 and a4). The 1.03~$\mu$m pump beam enters perpendicular to the front crystal plane (b1 and b2). The THz beam exits sideways as shown in Fig.~\ref{FIG1} (b3 and b4). (b2), (b4) Thermal images with optimized THz output. THz generation is minimized in (b1) and (b3).}
\label{FIG10}
\end{figure}

Fig.~\ref{FIG10} shows an example of the temperature distribution during THz generation in a \ce{LiNbO3} prism with Hebling design. Results from simulations (see below) appear in parts (a1)--(a4), and are compared to thermal images (model A655sc, FLIR) taken during operation in parts (b1)--(b4). Fig.~\ref{FIG10}(b1) and (b2) show a front view of the pump beam input facet. The heat signature of the absorbed pump beam and THz beam is visible close to the edge, where the pump beam enters the crystal. The pump beam is totally internally reflected on the THz beam exit facet (see Fig.~\ref{FIG10}(a4)), and is extracted on the opposite side, with a fraction of the beam being reflected towards the crystal apex. Remarkably, the heat signature of the pump and THz beam can be distinctly observed as the two beams follow separate optical paths. This is shown in Fig.~\ref{FIG10}(b2) and (b4) with optimized THz generation, and in Fig.~\ref{FIG10}(b1) and (b3), where THz generation is detuned. The maximum temperatures at the pump beam input (base, Fig.~\ref{FIG10}b1 and b2) are $66.97 \pm 0.24^{\circ}$C with THz on (b2) and $60.21 \pm 0.27 ^{\circ}$C with THz detuned (b1). The highest temperatures at the THz extraction point (Fig.~\ref{FIG10}(b3) and (b4)) are $67.74 \pm 0.44 ^{\circ}$C with THz on (b4) and $62.67 \pm 0.53 ^{\circ}$C with THz detuned (b3). The temperature at the crystal apex in Fig.~\ref{FIG10}(b3) and (b4) is $\sim 77^{\circ}$C. This measurement was performed with 162.3~W pump power. 

The \ce{LiNbO3} temperature distribution was modeled with finite element analysis in thermal steady state (Fig.~\ref{FIG10}(a1)--(a4)). The model parameters are thermal conductivities of 4.4~W/mK ($\parallel$) and 4.5~W/mK ($\perp$) (Ref.17 in~\cite{nikogosyan2006nonlinear}). The emissivity of the crystal, $\varepsilon = 0.71\pm0.01$, was measured using the high emissivity tape calibration method. The absorption coefficient of the pump wave is $\alpha_{1.03~\mu \mathrm{m}} = 1000~ppm/cm$ ~\cite{leidinger2015comparative}. The pump wave optical path inside the crystal is approximately 1.33~cm and the absorbed power is approximately $P_\text{abs} = 215$~mW. These values reproduce the results shown in Fig.~\ref{FIG10}(b1) and (b3) with no THz generation, within $\pm 0.5 ^{\circ}$C. Absorption of the THz wave causes additional temperature increase along the THz beam path as shown in Fig.~\ref{FIG10}(a2) and (a4). The optical path of the THz wave is approximately 0.35~cm. The modelled absorbed power of the THz wave is 37~mW, to reproduce temperatures measured in Fig.~\ref{FIG10}(b2) and (b4), within $\pm 1.2 ^{\circ}$C at the pump input and THz exit side.  

\begin{figure}[tb]
\centering\includegraphics[width=7.5cm]{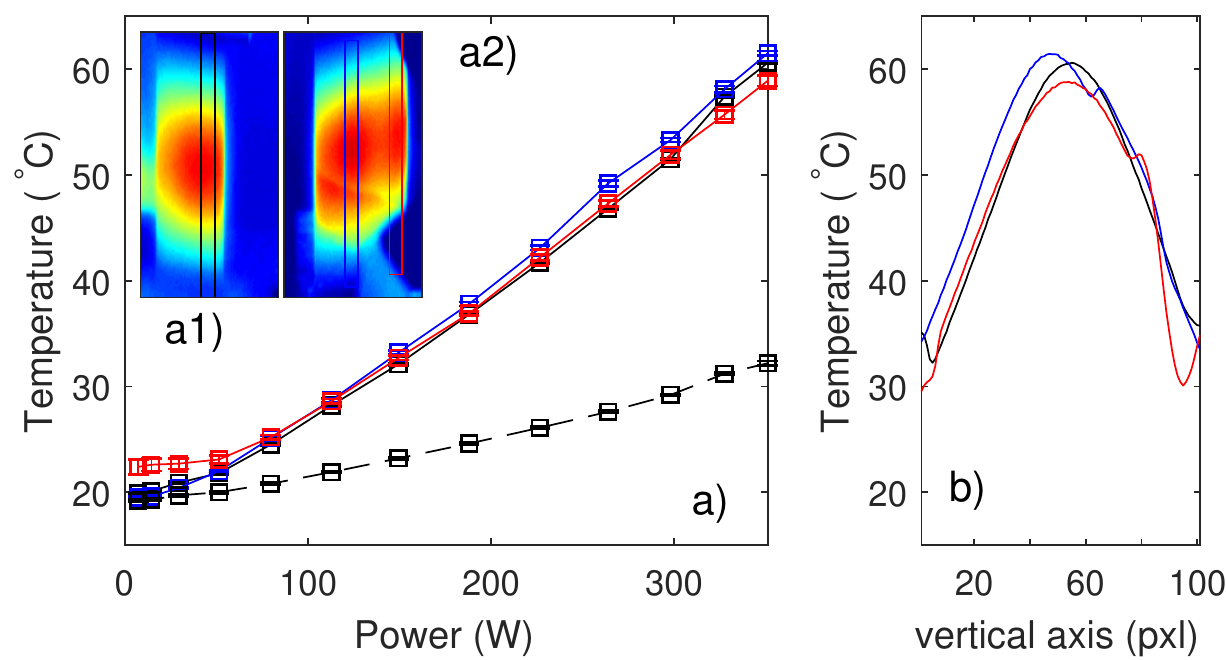}
\caption{Pump and THz wave absorption in \ce{LiNbO3}. (a) Power-dependent temperature measurements at three key locations: peak temperature at THz extraction location (blue), peak temperature at pump internal reflection location (red), and minimum temperature near cooling blocks (black). Insets: thermal images of (a1) pump input face and (a2) THz exit face at 351~W average pump power. (b) Mean lineout of black, blue and red rectangular areas marked in (a1) and (a2).}
\label{FIG11}
\end{figure}

The pump-power dependent peak temperatures and representative thermal gradients observed in the \ce{LiNbO3} crystal are displayed in Fig.~\ref{FIG11}. It is interesting to note that at both the THz beam exit spot and the region where the pump beam is internally reflected adjacent to it (see Fig.~\ref{FIG10}(b4) and Fig.~\ref{FIG11}(a2)), the peak temperatures, part (a), and gradients, part (b), are very similar. This suggests that heating of this section of the crystal is contributed to in almost equal parts by absorption of both THz and pump photons. Proper choice of the crystal geometry, with appropriate angles and having all three surfaces polished, has already mitigated much of the potential thermal impact from the incredibly high average power remaining in the pump beam. Both the isosceles- and Hebling-cut prisms were effective at coupling the vast majority of pump light safely into the beam dump (Fig.~\ref{FIG1}). A smaller \ce{LiNbO3} crystal with right-triangular prism geometry and only two sides polished was briefly investigated and showed thermal throttling of THz conversion at much lower pump powers (not shown), reinforcing crystal geometry as a key design choice. The thermal imaging study, however, shows that local heating on the THz output path from direct THz absorption is also a significant contributor that could only be mitigated by lessening the absorption or more efficiently removing heat. 

\section{Concluding remarks}
 We have demonstrated the generation of single-cycle THz pulses with energy up to 1.44~$\mu$J  at 100~kHz repetition rate. The peak THz electric field strength in the focus is $\sim 150$~kV/cm at a center frequency of 0.6~THz, with a maximum photon conversion efficiency of 19\% or energy conversion efficiency of $4.2 \times 10^{-4}$. Our results are enabled by a unique pump laser source based on multi-pass spectral broadening and recompression delivering sub-100~fs pulses with close to 1~kW average power. Using this laser source we were able to probe the limits of laser-driven THz generation through optical rectification in \ce{LiNbO3} crystals at high sustained pump power levels.  While the  crystals showed no signs of damage despite the very high pump intensity and average power, further scaling to even higher THz pulse energies and electric field values is limited primarily by thermally-induced absorption of the generated THz. We find that thermal management and efficient out-coupling of the depleted pump light from the crystal is crucial to achieve consistently large conversion efficiencies. We further notice that a significant part of the thermal load is caused by THz absorption within the \ce{LiNbO3} crystal. Hence, cryogenic cooling, which greatly reduces the absorption of THz radiation in \ce{LiNbO3}~\cite{hebling2004,Huang:13}, is expected to both increase the THz conversion efficiency in the pumping range explored presently, and allow scaling to even higher pump powers. We have demonstrated that a high energy, high repetition rate, multi-pass spectrally broadened pump can be an ideal source for enabling high-field THz excitation for next-generation high-repetition rate x-ray free electron laser experiments.

\section*{Appendix A: Pump laser parameters for THz experiments}

\begin{figure}[tb]
\centering\includegraphics[width=12.5cm]{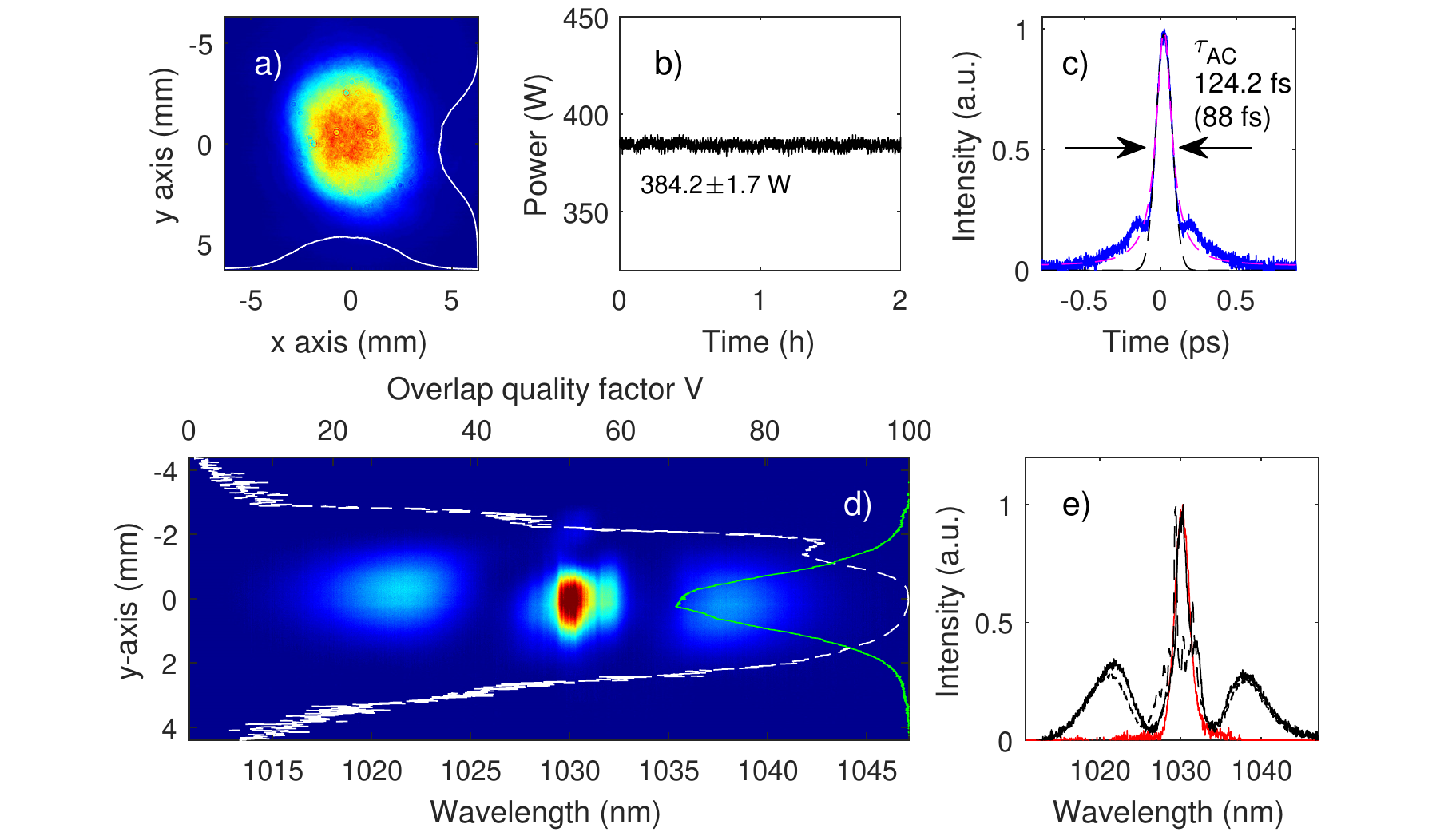}
\caption{Measurements of MPC output at 1.6~bar (used for THz generation). (a) Beam profile measured and (b) power measurement performed at the THz setup (black). (c) Pulse duration measurement (autocorrelation, black) and Gaussian fit ($\tau_{p}$ = 88~fs assuming a Gaussian temporal distribution, blue). Additional Lorentzian-fit (magenta) for comparison. (d) Spatio-spectral distribution of the broadened pulses. Overlap quality factor (white), integrated intensity distribution (green). (e) Integrated spectra for comparison (fundamental spectrum, red; 1.6~bar pressure, solid black; 1.1~bar pressure, dashed black).}
\label{FIG7:params}
\end{figure}

The Herriott cell is operated with a fixed beam size and compressor dispersion. The setup can be tuned using gas pressure $P_\text{gas}$ or peak power $P_\text{peak}$. For the THz generation experiments, the MPC was configured at at 1.6~bar and 7.12~GW peak power (compare with 1.1~bar and about 10~GW for the highest output power). Measurements of the operating conditions are shown in Fig.~\ref{FIG7:params}. The beam profile at the input of the THz setup is shown in Fig.~\ref{FIG7:params}(a). The power measurement shown in Fig.~\ref{FIG7:params}(b) is the highest output used in the THz setup. The pulse duration (Fig.~\ref{FIG7:params}(c)) is 88~fs assuming a Gaussian temporal distribution. The spatio-spectral distribution at 1.6~bar is shown in Fig.~\ref{FIG7:params}(d). The integrated spectra in two spectral broadening configurations, 1.1~bar (dashed black) and 1.6~bar (solid black), along with the fundamental spectrum (red) are shown in Fig.~\ref{FIG7:params}(e) for comparison. 

\section*{Appendix B: THz conversion efficiency estimation}

\begin{figure}[htb]
    \centering
    \includegraphics[width=8cm]{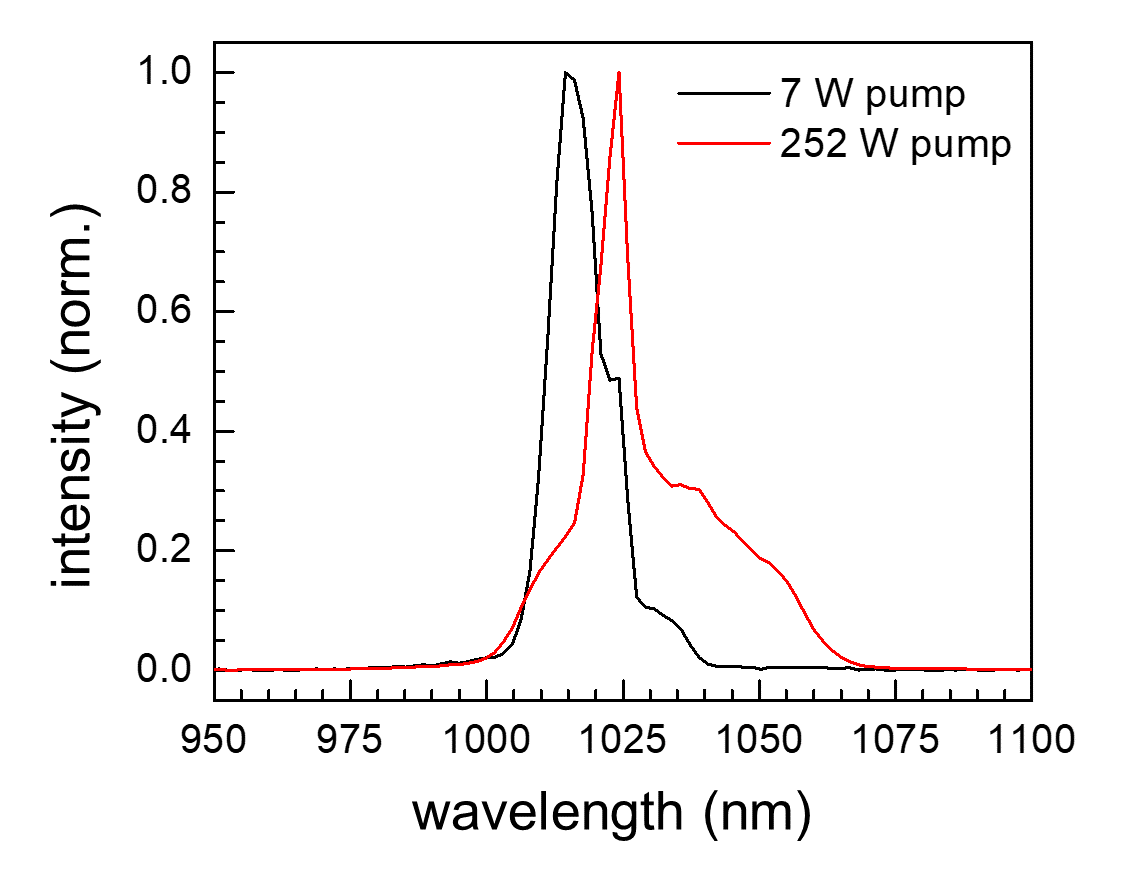}
    \caption{Spectra of residual pump light coupled out of \ce{LiNbO3} prism at low pump power with minimal THz generation (black) and at relatively high power with efficient THz generation (red). Both a shift of the detected peak frequency towards the red and an extended shoulder on the long-wavelength side are apparent when there is strong THz output. The shoulder is due to cascaded red-shifting of pump photons through the optical rectification process.}
    \label{FigTHzS2}
\end{figure}

Fig.~\ref{FigTHzS2} displays the spectra of the residual pump light coupled out the back face of the \ce{LiNbO3} prism at both low (7~W) and high (252~W) input power, directed into a spectrometer (Ocean Optics) by a fiber optic cable with the input anchored adjacent to the beam dump. The pump beam exiting the \ce{LiNbO3} prism (Hebling geometry) has angular dispersion, originating both from the imparted pulse front tilt as well as refraction out of the prism at non-normal incidence. Thus the spectra displayed in Fig.~\ref{FigTHzS2}, recorded through coupling into the fiber from a specific section of the beam dump, do not actually capture the entire pump spectrum. The overall shift in the peak between the low- and high-power spectra can most likely be attributed to the significant change in temperature of the nonlinear crystal, increasing from room temperature to over $60^\circ$C in the center of the pumped spot. The temperature thus changes the index of refraction and therefore the dispersion of the prism, making a small but noticeable change to the detected range of frequencies by the fiber-coupled spectrometer in a fixed location.

The most informative part of the pump spectra shown in Fig.~\ref{FigTHzS2} is the clear broadened shoulder on the long-wavelength edge when there is considerable THz generation, corresponding to a high pump pulse energy. To analyze purely the effect this shoulder has on the distribution of pump frequencies, the higher-power pump spectrum was shifted in frequency so that the peak value occurred at the same frequency position as for the low power spectrum. Once this thermally-induced change to the detected pump spectra was removed, any remaining change in the frequency distribution could be attributed to the THz generation process by optical rectification (OR). 

Mean frequencies (in THz units) from the intensity spectra $I_k(\omega)$ are calculated as
\[ 
\left\langle \omega_k \right\rangle = \frac{\int_0^\infty \mathrm{d} \omega \, \omega I_k(\omega) }{\int_0^\infty \mathrm{d} \omega \, I_k(\omega)}, 
\]
with $k$ taking on the values of: $I$ for the initial, undepleted, pump at low THz conversion, $F$ for the final, broadened and red-shifted, pump coinciding with efficient THz generation, and $T$ for the resulting intensity-level THz spectrum. Generation of a THz photon, with average frequency $\left\langle \omega_T \right\rangle$, results in generation of a pump photon which is redshifted by $\left\langle \omega_T \right\rangle$ from the average input frequency, $\left\langle \omega_I \right\rangle$, satisfying energy conservation. Each additional step of OR induces one further redshift by the THz frequency while generating an additional THz photon. Thus we can calculate the photon-level conversion efficiency as~\cite{Yeh_2007}
\[
\eta_\text{photon} = \frac{\left\langle \omega_I \right\rangle - \left\langle \omega_F \right\rangle}{\left\langle \omega_T \right\rangle},
\]
which gives the average number of THz photons generated by each incoming pump photon. Because OR can be cascaded, with each pump photon undergoing multiple red-shifts and thus generating multiple THz photons, $\eta_\text{photon}$ can be greater than unity. 
For calculation of the energy conversion efficiency, we must account for the ratio of THz and pump photon energies (or frequencies):
\[
\eta_\text{energy} = \frac{\left\langle \omega_T \right\rangle}{\left\langle \omega_I \right\rangle} \times \eta_\text{photon} = \frac{\left\langle \omega_I \right\rangle - \left\langle \omega_F \right\rangle}{\left\langle \omega_I \right\rangle}.
\]
The energy efficiency is always less than one. 

For the pump spectra shown in Fig.~\ref{FigTHzS2}, we find an average pump redshift of 1.493~THz. The THz intensity spectrum (see field distributions in Fig.~\ref{FigTHz1}(e) and Fig.~\ref{FigTHzS3}) gives a mean terahertz photon frequency of 0.607~THz. Thus, internal to the \ce{LiNbO3} crystal, the conversion efficiencies are $\eta_\text{photon} = 2.46$ and $\eta_\text{energy} = 5.1 \times 10^{-3}$. Considering the high index of the lithium niobate crystal for THz, $n_\text{THz} =  4.96$, the energy transmission factor for normal incidence out-coupling is 0.5585. Therefore, the ``external'' conversion efficiencies are $\eta_\text{photon, ext.} = 1.37$ and $\eta_\text{energy, ext.} = 2.8 \times 10^{-3}$.

\section*{Funding}

Use of the Linac Coherent Light Source, SLAC National Accelerator Laboratory, is supported by the US Department of Energy, Office of Science, Office of Basic Energy Sciences, under contract no. DE-AC02-76SF00515. M.C.H. is supported by the US Department of Energy, Office of Science, Office of Basic Energy Sciences, under award no. 2015-SLAC-100238-Funding.

\section*{Acknowledgments}

The authors would like to thank Torsten Mans, Arvid Hage and other team members of Amphos GmbH, for their support and for insightful discussions.

\section*{Disclosures}

The authors declare no conflicts of interest.





\bibliography{SBandTHz}

\end{document}